\documentclass[pra,aps,twocolumn,showpacs,amsmath,amssymb,floatfix,10pt]{revtex4-1}
\input{epsf}

\usepackage{graphicx}
\usepackage{dcolumn}
\usepackage{bm}

\begin{document}
\title{Quantum phase measurement and Gauss sum factorization of large integers in a superconducting circuit}
\author{H. T. Ng${}^{1}$ and Franco Nori${}^{1,2}$}
\affiliation{${}^{1}$Advanced Science Institute, RIKEN, Wako-shi 351-0198, Japan}
\affiliation{${}^{2}$Physics Department, The University of Michigan, Ann Arbor, Michigan 48109-1040, USA}
\date{\today}

\begin{abstract}
We study the implementation of quantum phase measurement in a superconducting circuit, where 
two Josephson phase qubits are coupled to the photon field inside a resonator. 
We show that the relative phase of the superposition of two Fock states can be 
imprinted in one of the qubits.  The qubit can thus be used to probe and store the quantum coherence of two distinguishable 
Fock states of the single-mode photon field inside the resonator.  The effects of dissipation of the photon 
field on the phase detection are investigated.
We find that the visibilities can be greatly enhanced if the Kerr nonlinearity is exploited.
We also show that the phase measurement method can be used to perform the Gauss sum factorization of 
numbers (${\geq}~10^4$) into a product of prime integers, as well as to precisely measure both the resonator's frequency
and the nonlinear interaction strength.  
The largest factorizable number is mainly limited by the coherence time.
If the relaxation time of the resonator were to be ${\sim}~10$ $\mu$s~(${\sim}~1$ ms), 
then the largest factorizable number can be ${\geq}~10^4N$~(${\geq}~10^{7}N$),
where $N$ is the number of photons in the resonator.
\end{abstract}


\maketitle

\section{Introduction}
The superposition of states is a fundamental feature of quantum mechanics.
Recently, the arbitrary superposition of Fock states \cite{Law,Liu} has been produced in a superconducting
resonator with a Josephson phase qubit \cite{Hofheinz1}.  This offers novel ways 
to directly study the quantum coherence of the photon field, i.e., superposition of number states.  
Also, this strongly coupled qubit-resonator system \cite{Hofheinz2,Wang}, may be useful 
for quantum information processing (QIP).

\subsection{Quantum phase measurement}
We theoretically study the quantum phase measurement of the photon field in
a superconducting resonator coupled to two phase qubits. 
We consider probing the quantum coherence of the superposition state
\begin{eqnarray}
\label{multi-photon}
|\Psi_N\rangle&=&\frac{1}{\sqrt{2}}\big[|0\rangle+\exp{(i\varphi_N)}|N\rangle\big],
\end{eqnarray}
where $|0\rangle$ and $|N\rangle$ are the vacuum and the multi-photon state, respectively, and $\varphi_N$
is the relative phase.  This superposition of states leads to interference fringes.
However, in a dissipative environment,
the quantum coherence of the superposition in Eq.~(\ref{multi-photon}) 
decays rapidly as $N$ grows large \cite{Walls}.  Systematically studying such superpositions should provide a better
understanding of the decoherence process \cite{Walls,Wang2}.

To measure the quantum state of the photon field, Wigner tomography can be used \cite{Hofheinz1,Bertet}.  The relative phase
between two number states can only cause a rotation in Wigner phase space without changing the shape of the Wigner function 
\cite{Hofheinz1}. 
Alternatively, here we propose a method to transfer the phase information of the photon field to
the qubit, such that we can determine the relative phase precisely by measuring the quantum state of the qubit.
In this way, the qubit can be used to store and detect the quantum coherence of
an arbitrary superposition of two photon number states.  

The superposition of the vacuum and the single-photon state $|\Psi_1\rangle$ 
for $N=1$ in Eq.~(\ref{multi-photon}) in a microwave cavity 
has been used for the quantum memory of an atomic qubit \cite{Maitre}.   
This experiment has demonstrated that the quantum information of the qubit can be transferred 
to the photon field.
However, here we find that it is necessary to use two qubits to transfer the phase information
of the superposition of the vacuum and the multi-photon state $|\Psi_N\rangle$ in Eq.~(\ref{multi-photon}).  The first qubit is used for 
storing the quantum information of the photon field, whereas the second qubit is used
as an auxiliary qubit to disentangle the first qubit from the resonator \cite{Barenco} by repeatedly applying
a controlled-NOT (CNOT) quantum gate (see e.g., \cite{You1,You2,Yamamoto,Plantenberg}).  
Several proposals (e.g., \cite{Galiautdinov,Geller}) have been made to implement CNOT gates using superconducting qubits.

Our proposed quantum phase detection method can measure the degree of quantum coherence of the photon field, 
which can be determined by the visibility of the detection signal \cite{Walls}.  However, the visibilities are
greatly reduced due to decoherence, and depend on the quality factor \cite{Liu2} (i.e., ratio of the frequency and
the damping rate of the resonator).  We find that {\it the visibilities can be greatly enhanced if 
the Kerr nonlinearity of the cavity mode} (e.g., \cite{Gong,Semiao}) {\it is exploited}.  
Thus, the interference of the superposition of multi-photon states can be observed 
clearly even in the presence of a dissipative environment.
Notably, the production of extremely strong Kerr nonlinear strength via coupled 
Cooper Pair Boxes (CPBs) \cite{You} using the effect of electromagnetically induced transparency (EIT) 
(e.g., \cite{Harris,Imamoglu,Scully}) 
has recently been proposed \cite{Ian,Rebic}.  The EIT effect in a lossless medium, such as quantum dots embedded in a solid-state
substrate, has recently been studied \cite{He}.

\subsection{Gauss sums}
The big problem with Shor's algorithm \cite{Shor,Wei} is that it is 
very difficult to implement physically for numbers larger than 15.  
Thus, here we use the Gauss sum approach \cite{Mack,Mehring,Gilowski,Sadgrove,Bigourd} because this is 
implementable and could provide a very valuable 
test-bed or stepping stone for more powerful future implementations.
Our proposed phase detection scheme can be used for implementing the Gauss sum, which can find the factors of a number
using the periodicity of the sum of the quadratic phase factors \cite{Mack}.   
The Gauss sum has been realized with NMR \cite{Mehring}, cold atoms \cite{Gilowski,Sadgrove} and using
short laser pulses \cite{Bigourd}.  Here we study the implementation of the Gauss sum in  
superconducting circuits.   Using our proposed superconducting circuits, factors of 
integers $\sim~{10^3}$ should be obtainable by the Gauss sum if the relaxation times were 
to be several $\mu$s.  The size of the factors are
mainly limited by the coherence time of the photon field 
in the superconducting resonator.  Thus, the factors of much larger numbers could be obtained
in the future, when coherence times improve.  

\subsection{Measuring the frequency}
We also show that the quantum phase measurement approach proposed here can be applied to precision measurements.
This approach enables to precisely determine the frequency of the resonator and the strength of 
the Kerr nonlinearity.  
We show that the superposition of multi-photon states can increase the accuracy
of precision measurements.  
This can act as a ``frequency standard'' for the other qubits in the circuit.

This paper is organized as follows: in Sec.~II, we describe the model of the system studied.
In Sec.~III, we present a method to detect the relative phase of the superposition of the
vacuum and the multi-photon state in Eq.~(\ref{multi-photon}).  In Sec.~IV, we show that the phase detection
method can be applied to both the Gauss sum for factorization and precision measurements.  We close the paper with
a summary.  In appendix A, we study the interaction between a qubit and a resonator in 
the far-detuning regime.  In appendix B, we discuss the effect of imperfect CNOT gate operations 
on the disentanglement process.

\section{System}
\begin{figure}[ht]
\centering
\includegraphics[height=5.8cm]{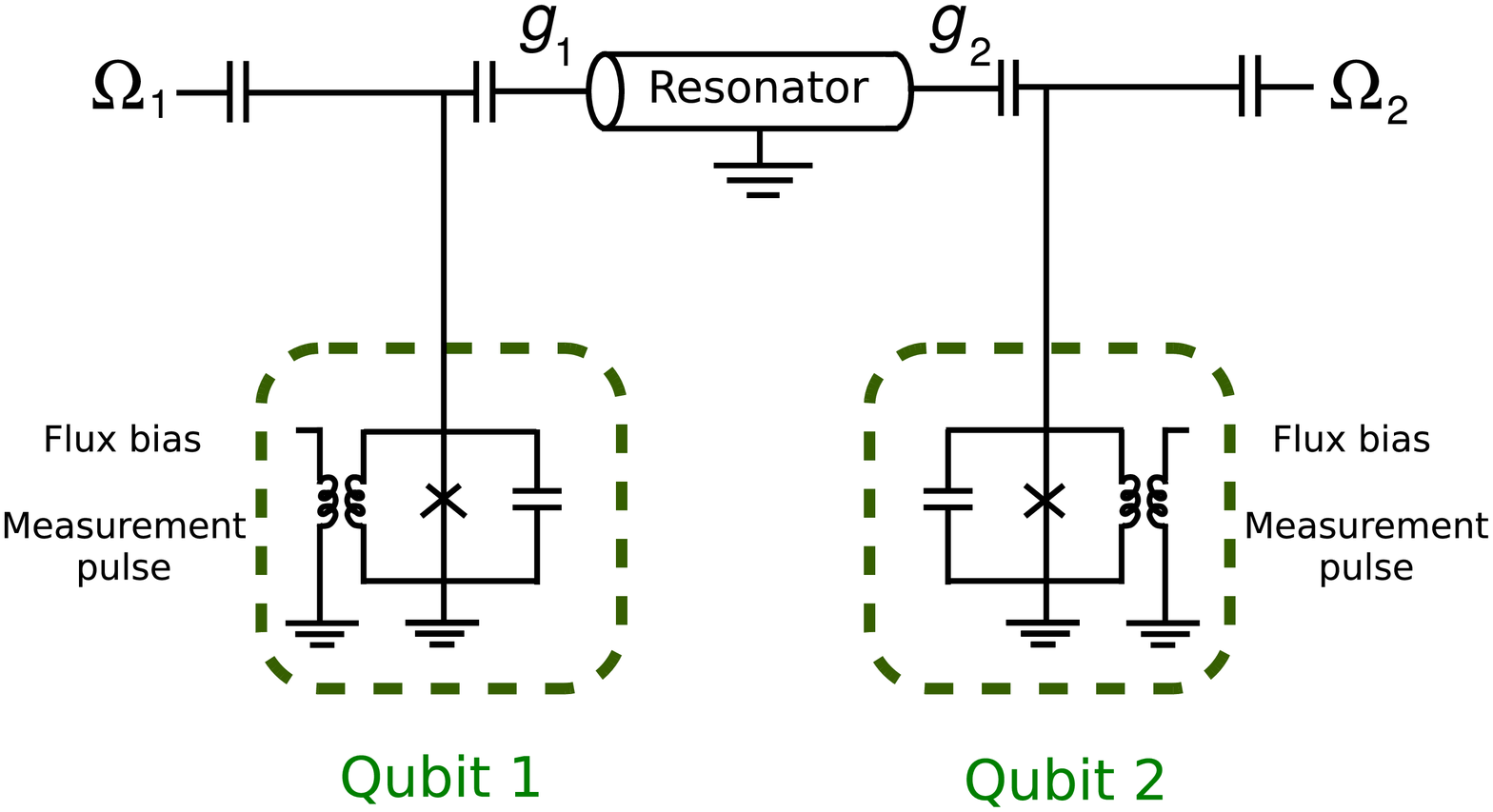}
\caption{ \label{fig1} (Color online) Circuit diagram of the qubits and the resonator for quantum phase measurements. The Josephson qubits
are capacitively coupled to the coplanar waveguide resonator with couplings $g_1$ and $g_2$.  The frequency of the qubit can be adjusted
by the flux bias pulse and can be measured by a superconducting quantum interference device (SQUID) \cite{Martinis}.
Here $\Omega_1$ and $\Omega_2$ are the amplitudes of the couplings between the ground and excited states.}
\end{figure}

We consider two Josephson phase qubits capacitively coupled to a superconducting resonator \cite{Martinis}
as shown in Fig.~\ref{fig1}.  
The Hamiltonian of the qubit-resonator system can be written as \cite{Hofheinz1} $(\hbar=1)$ 
\begin{eqnarray}
\label{Ham1}
H\!&=&\!H_{\rm res}+\sum^{2}_{j=1}\big[H^{(j)}_{\rm qbit}+H^{(j)}_{\rm qbit-res}+H^{(j)}_{\rm drive}\big],\\
&=&\omega{a^\dag{a}}+\sum^2_{j=1}\Big\{\frac{\omega_{0j}}{2}\sigma_{jz}+g_j(a\sigma_{j+}+\sigma_{j-}a^\dag)\nonumber\\
&&\!\!-\frac{\Omega_j}{2}\big[\exp{(-i\phi_j-i\omega_{qj}{t})}\sigma_{j+}+{\rm H.c.}\big]\Big\},
\end{eqnarray}
where $a$ is the annihilation operator of the photon field, $\sigma_{j\pm}$ and $\sigma_{jz}$
are the transition and population (Pauli $z$) operators of the qubit $j$ respectively ($j=1,2$).  
Here $\omega_{0j}$ and $\omega$ are the frequencies of the qubit $j$ and the resonator, respectively.  
The parameters $g_j$ act as coupling strengths between the photon field and the qubits.
Also, $\Omega_j$ and $\phi_j$ are the amplitude and the phase of the coupling between the ground state $|g\rangle$ and 
the excited state $|e\rangle$ in the $j$-th qubit.  The frequency $\omega_{qj}$ is the frequency of 
the microwave drive of the qubit $j$.
The qubit $j$ can be accurately controlled by adjusting the frequency $\omega_{0j}$ and the parameter
$\Omega_j$ via a classical signal \cite{Hofheinz1,Hofheinz2}.  

It is convenient to work in the interaction picture.  By considering the unitary transformation
\begin{equation}
\label{U}
U(t)=\exp\Big[-i\Big(\omega{a^\dag{a}}+\frac{\omega_{01}}{2}\sigma_{1z}+\frac{\omega_{02}}{2}\sigma_{2z}\Big)t\Big], 
\end{equation}
the transformed Hamiltonians are
\begin{eqnarray}
H^{(j)}_1&=&U^\dag(t)H^{(j)}_{\rm drive}U(t),\\
&=&-\frac{\Omega_j}{2}\big[\exp{(-i\phi_j+i\tilde{\Delta}_j{t})}\sigma_{j+}+{\rm H.c.}\big],
\end{eqnarray}
and
\begin{eqnarray}
H^{(j)}_2&=&U^\dag(t)H^{(j)}_{\rm qbit-res}U(t),\\
&=&g_j\big[\exp(i\Delta_j{t})a\sigma_{j+}+{\rm H.c.}\big],
\end{eqnarray}
where 
\begin{eqnarray}
\tilde{\Delta}_j&=&\omega_{0j}-\omega_{qj},
\end{eqnarray}
and
\begin{eqnarray}
\Delta_j&=&\omega_{0j}-\omega.
\end{eqnarray}
Hereafter, the two transformed Hamiltonians $H^{(j)}_{1}$ and $H^{(j)}_{2}$ at resonance (i.e., when $\Delta_j=0$ and $\tilde{\Delta}_j=0$), 
\begin{eqnarray}
\label{qubit_drive}
H^{(j)}_{1}&=&-\frac{\Omega_j}{2}\big[\exp{(-i\phi_j)}\sigma_{j+}+{\rm H.c.}\big],\\
\label{qubit_resonator}
H^{(j)}_{2}&=&g_j(a\sigma_{j+}+\sigma_{j-}a^\dag),
\end{eqnarray}
will be used frequently.
Also, the time evolution operators $U^{(j)}_1(t)=\exp[-iH^{(j)}_{1}t]$ and $U^{(j)}_2(t)=\exp[-iH^{(j)}_{2}t]$ can be obtained explicitly as \cite{Scully}
\begin{eqnarray}
\label{U1}
U^{(j)}_1(t)&=&\cos\bigg(\frac{\Omega_j{t}}{2}\bigg)\openone+i\sin\bigg(\frac{\Omega_j{t}}{2}\bigg)\big[\exp{(-i\phi_j)}\sigma_{j+}\nonumber\\
&&+{\rm H.c.}\big],
\end{eqnarray}
and
\begin{eqnarray}
\label{U2}
U^{(j)}_2(t)&=&\cos\big(g_jt\sqrt{aa^\dag}\big)|e\rangle\langle{e}|+\cos\big(g_jt\sqrt{a^\dag{a}}\big)|g\rangle\langle{g}|\nonumber\\
&&-ia\frac{\sin{(g_jt\sqrt{a^\dag{a}})}}{\sqrt{{a^\dag}{a}}}\sigma_{j+}-ia^\dag\frac{\sin(g_jt\sqrt{aa^\dag})}{\sqrt{aa^\dag}}\sigma_{j-},\nonumber\\
\end{eqnarray}
where $\openone$ is the unit operator.

To transfer the phase information of the photon field to the qubit, it is required to switch-on and off the
interaction between the qubit and the resonator.
The coupling between the qubit and the resonator is fixed, but the frequency of the qubit can be adjusted by the bias
current \cite{Hofheinz1,Hofheinz2}.  The qubit-resonator interaction can thus be turned-off by far-tuning 
the frequency $\omega_{0j}$ of the phase qubit \cite{Hofheinz1,Hofheinz2}.  In appendix A, we give a detailed discussion of the 
qubit-resonator coupling in the far-detuning regime.

\section{Quantum phase measurement}
We now present a procedure to use the qubits to probe and 
store the quantum coherence of the superposition of two Fock states.
The relative phase between the superposition of the vacuum and the {\it single}-photon state
in Eq.~(\ref{multi-photon})
can be completely transferred to the qubit such that the qubit can act as a probe of 
the quantum coherence of the photon field \cite{Maitre}.  The phase can be determined 
by measuring the qubit's state.

We require two qubits to measure the phase of the superposition state of
the vacuum and the {\it multi}-photon state  in Eq.~(\ref{multi-photon}).  
One qubit is used for probing the phase and the other qubit
is used to disentangle the qubit from the resonator.  In the following subsections, 
we will describe the different schemes for the single- and multi-photon cases.

\subsection{Single-photon case: $|0\rangle+\exp(i\varphi_1)|1\rangle$}
Let us now consider the quantum phase measurement of the superposition, $[|0\rangle+\exp(i\varphi_1)|1\rangle]/\sqrt{2}$, 
of the vacuum and the single-photon Fock state
in the resonator.  We can create the superposition state $|\Psi_1\rangle$ in Eq.~(\ref{multi-photon}) 
by just using one qubit \cite{Liu,Hofheinz1,Maitre}.  
Initially, the product state of the vacuum of the resonator and the ground state of the qubit 1
is prepared, i.e., $|{\rm qubit},{\rm resonator}\rangle=|g,0\rangle$. 
We first produce an equal superposition of the states $|g\rangle$ and $|e\rangle$
of the qubit 1 by applying a $\pi/2$-pulse to the qubit 1 (i.e., turning on the drive of the qubit 1 for a
time $T=\pi/2\Omega_1$).  By applying the time-evolution
operators $U(t)U^{(1)}_1(t)$ in Eqs. (\ref{U}) and (\ref{U1}) to the state $|g\rangle$, we have
\begin{eqnarray}
&&U(t)U^{(1)}_1(t)|g\rangle\nonumber\\
&=&U(t)\Big[\cos\Big(\frac{\Omega_1{t}}{2}\Big)|g\rangle+i\exp(-i\phi_1)\sin\Big(\frac{\Omega_1{t}}{2}\Big)|e\rangle\Big].\nonumber\\
\end{eqnarray}
Now consider $U^{(1)}_1(t=T=\pi/2\Omega_1)$, then
\begin{eqnarray}
&=&\frac{1}{\sqrt{2}}U(t)\big[|g\rangle+i\exp(-i\phi_1)|e\rangle\big],\nonumber\\
&=&\frac{1}{\sqrt{2}}\exp\Big(\frac{i\omega_{01}t}{2}\Big)\big[|g\rangle+i\exp{(-i\phi_1')}|e\rangle\big],
\end{eqnarray}
where $\phi_1'=\phi_1+\omega_{01}t$.
The states can thus be written as
\begin{eqnarray}
|\Phi_1(0)\rangle&=&U(t)U^{(1)}_1(t=\pi/2\Omega_1)|g,0\rangle,\\
\label{Phi_1}
&=&\frac{1}{\sqrt{2}}\exp\Big(\frac{i\omega_{01}t}{2}\Big)\big[|g\rangle+i\exp{(-i\phi_1')}|e\rangle\big]|0\rangle.\nonumber\\
\end{eqnarray}
Now for $U(t=\pi/2\Omega_1)$, then 
\begin{eqnarray}
\label{pi/2}
|\Phi_1(0)\rangle&=&\frac{1}{\sqrt{2}}\exp\Big(\frac{i\omega_{01}\pi}{4\Omega_1}\Big)\big[|g\rangle+i\exp{(-i\phi_1')}|e\rangle\big]|0\rangle.\nonumber\\
\end{eqnarray}

Next, we turn on the qubit-resonator interaction for a time $t^*_1$,
\begin{equation}
\label{timeqr}
t^*_{1}=\frac{\pi}{2g_1}.
\end{equation}
The energy of excited state of qubit 1 then lowers one level, to its ground state, and a photon 
is created in the resonator.  This can be derived by applying the evolution operator $U(t^*_1)\,U^{(1)}_2\!(t^*_1)$ in Eqs. 
(\ref{U},\ref{U2}) giving
\begin{eqnarray}
\label{g1}
U(t^*_1)\,U^{(1)}_2\!(t^*_1)|e,0\rangle&=&-i\exp(-i\omega{t^*_1})|g,1\rangle,
\end{eqnarray}
and the ground state $|g\rangle|0\rangle$ has not changed,
\begin{eqnarray}
\label{g0}
U(t^*_1)\,U^{(1)}_2\!(t^*_1)|g,0\rangle&=&|g,0\rangle.
\end{eqnarray}
Combining Eqs. (\ref{Phi_1},\ref{g1},\ref{g0}),  we obtain the state
\begin{eqnarray}
\label{singlestate}
|\Phi_1(t^*_1)\rangle&=&U(t^*_1)\,U^{(1)}_2\!(t^*_1)|\Phi_1(0)\rangle,\\
&=&\frac{1}{\sqrt{2}}|g\rangle\big\{|0\rangle+\exp{[-i(\phi_1'+\omega{t^*_1})]}|1\rangle\big\}.
\end{eqnarray}
Here we have ignored the global phase factor $\exp(i\omega_{01}\pi/4\Omega_1)$ in Eq.~(\ref{pi/2}).

We then switch-off the qubit-resonator interaction and 
let the system evolve freely for a short period $\tau$, 
such that a relative phase is acquired 
between the two number states $|0\rangle$ and $|1\rangle$.  
The total state, at the time $t'=t^*_1+\tau$, becomes
\begin{eqnarray}
|\Phi_1(t')\rangle&=&\exp(-i\omega{a}^\dag{a}\tau)|\Phi_1(t^*_1)\rangle,\\
&=&\frac{1}{\sqrt{2}}|g\rangle\big\{|0\rangle+\exp{[-i\phi_1'-i\omega{(t^*_1+\tau)}]}|1\rangle\big\}.\nonumber\\
\end{eqnarray}

We can now transfer the phase information to the qubit 1 by 
switching-on the qubit-resonator interaction for the time $t^*_1$ in Eq. (\ref{timeqr}). 
The state $|g\rangle|1\rangle$ will swap to $|e\rangle|0\rangle$, while the ground
state $|g\rangle|0\rangle$ remains unchanged.  The state now reads
\begin{equation}
\label{proqrs}
|\Phi_1(t)\rangle=\frac{1}{\sqrt{2}}\big\{|g\rangle-i\exp[-i(\phi_1'+\varphi_1)]|e\rangle\big\}|0\rangle,
\end{equation}
where $t=2t^*_1+\tau$ and $\varphi_1=\omega{t}\approx\omega{\tau}$.  
Here we have assumed that the period $\tau$ is much greater than the time $t^*_1$.
The relative phase information between the two number states is now imprinted in the qubit 1
and the qubit 1 is disentangled from the photon field in the superconducting resonator.

Let us apply a $\pi/2$-pulse to the qubit 1 so that the final state can be written as
\begin{eqnarray}
|\Phi_1(t_f)\rangle&\approx&\frac{1}{2}\exp\Big(\frac{i\omega_{01}T}{2}\Big)\Big\{[1+\exp(-i\varphi_1')]|g\rangle\nonumber\\
&&+{i\exp(-i\phi_1')}[1-\exp(-i\varphi_1')]|e\rangle\Big\}|0\rangle,
\end{eqnarray}
where $t_f=t+T$, $\varphi_1'=\varphi_1-\omega_{01}T/2$ and $T=\pi/2\Omega_1$.  We have also assumed that the time 
$T=\pi/2\Omega_1$ is much shorter than the time $t$ and we approximate $\varphi_1'{~\approx~}\varphi_1$.
The excited state of the qubit 1 (with a much higher tunneling
rate than that of the ground state) can now be measured.  This can be done by applying a measurement pulse and 
read-out by a SQUID \cite{Martinis}.
The phase factor can thus be determined from the probability of the excited state, which is
\begin{eqnarray}
\label{prob_e}
P_e&=&\frac{1}{2}(1-\cos\varphi_1).
\end{eqnarray}

\subsubsection{Dissipation in the photon field}
The photon field inevitably suffers from the dissipation present
in realistic situations.  The thermal average photon number is about zero ($\sim{10^{-6}}$) for a high-frequency 
resonator ($\sim~{40}~$GHz) at low temperatures ($\sim$ ${25}$ mK) \cite{Hofheinz1}. 
The time evolution of the density matrix $\rho$ of the photon 
field can be described by the master equation \cite{Barnett}
\begin{eqnarray}
\label{master}
\dot{\rho}&=&-i{\omega}[a^\dag{a},\rho]+\Gamma(2a\rho{a^\dag}-a^\dag{a}\rho-{\rho}a^\dag{a}),
\end{eqnarray}
where $\Gamma$ is the damping rate.  We assume that the number of thermal photons is negligible.
Here we ignore the decoherence effect during
the qubit and qubit-resonator operations because their time durations ($T$ and $t^*$ $\sim~$ns in \cite{Hofheinz1}) 
are extremely short compared to the dissipation time $\Gamma^{-1}$ (several $\mu$s in \cite{Hofheinz1}).
Therefore, we consider the dissipation of the photon field during the free time-evolution $\tau$. 
The density matrix $\rho$ of the photon field at the time $\tau$ can then be found as \cite{Barnett}
\begin{eqnarray}
\rho&=&\frac{1}{2}\big\{2|0\rangle\langle{0}|-\exp({-2\Gamma{\tau}})(|0\rangle\langle{0}|-|1\rangle\langle{1}|)\nonumber\\
&&+\exp(-\Gamma{\tau})[\exp(i\omega{\tau}+i\phi_1')|0\rangle\langle{1}|+{\rm H.c.}]\big\}.
\end{eqnarray}
We then follow the same procedures discussed above.
The probability $P_e$ that the qubit is in its excited state can be readily obtained 
\begin{eqnarray}
\label{P_es}
{P}_e&=&\frac{1}{2}[1-\exp({-\Gamma{\tau}})\cos\varphi_1],
\end{eqnarray}
where $\varphi_1=\omega\tau$.

\subsubsection{Visibility}
The visibility of the quantum coherence can be defined as 
\begin{eqnarray}
\label{visibility}
V&=&\frac{C_{\rm max}-C_{\rm min}}{2},
\end{eqnarray}
where $C_{\rm max}$ and $C_{\rm min}$ denote the maximum and minimum values of the coherence
factor $C(\tau)$, 
\begin{equation}
C(\tau)=\exp{(-\Gamma{\tau})}\cos\varphi_1,
\end{equation}
which characterizes the quantum coherence of the superposition state.   Note that the visibility $V$ is unity
when dissipation is absent.
The visibility $V$ can be obtained as 
\begin{eqnarray}
V&=&\frac{1}{2}\bigg[1+\bigg(\frac{\omega^2}{\Gamma^2+\omega^2}\bigg)^{\frac{1}{2}}\exp{(-\Gamma{\tau_m})}\bigg],
\end{eqnarray}
where 
\begin{equation}
\tau_m=\frac{1}{\omega}\arccos\bigg[-\bigg(1+\frac{\Gamma^2}{\omega^2}\bigg)^{-\frac{1}{2}}\bigg],
\end{equation}
and $\pi/2~{\leq}~\omega\tau_m~{\leq}~\pi$.
Clearly, a higher visibility can be obtained for larger resonator quality factors 
(i.e., ratios of $\omega$ and $\Gamma$).

\subsection{Multi-photon case: $|0\rangle+\exp(i\varphi_N)|N\rangle$}
\begin{figure}[ht]
\centering
\includegraphics[height=5.8cm]{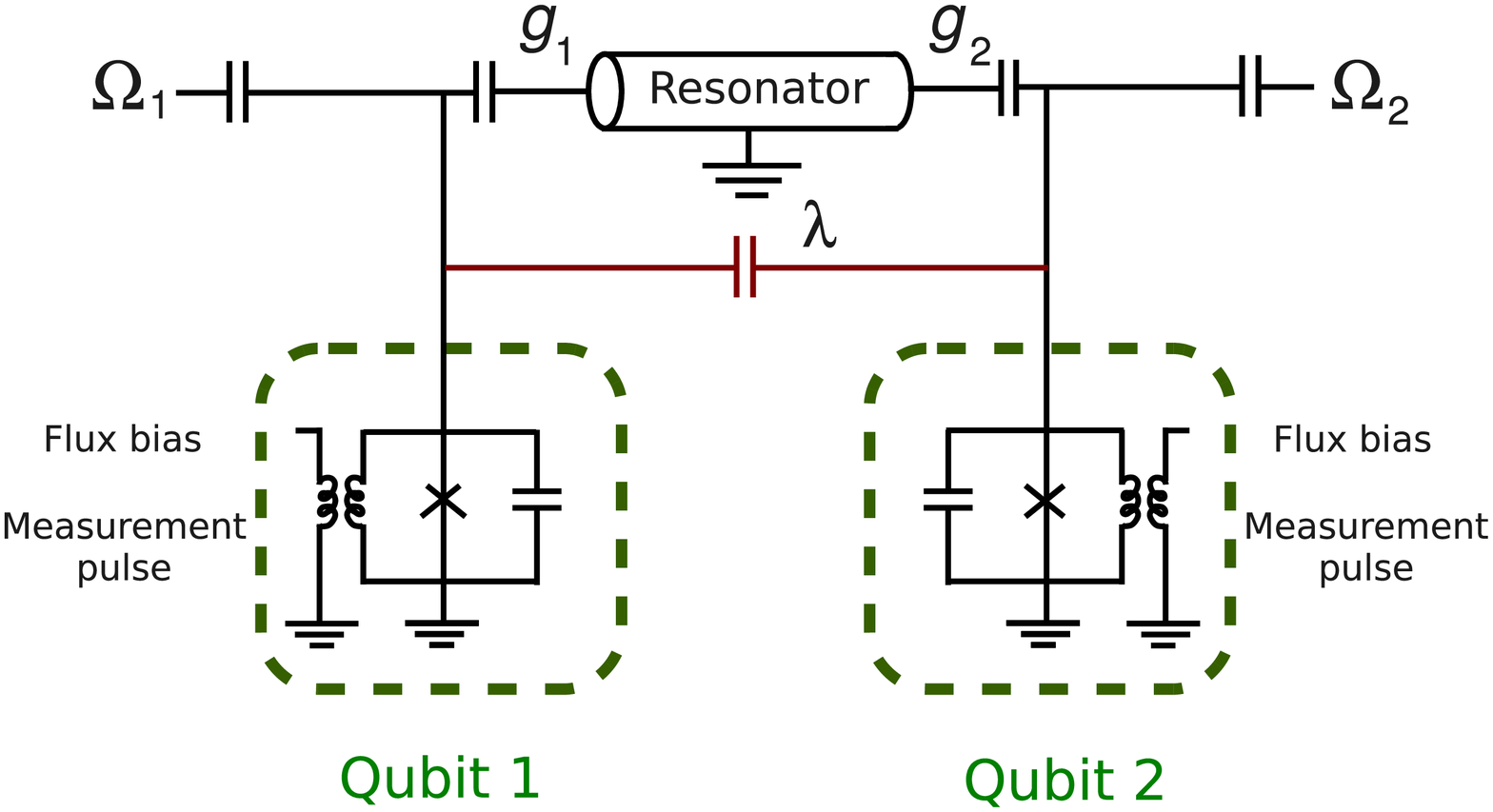}
\caption{ \label{fig2} (Color online) Circuit diagram of the coupled qubits and the resonator. The Josephson qubits
can be coupled to each other via a capacitor (shown in dark red line).  Here $\lambda$ is the coupling strength between
the two qubits.}
\end{figure}

We now investigate the quantum phase detection of the superposition of the vacuum and
multi-photon states in Eq. (\ref{multi-photon}), where $N{~\geq~}2$.  
This superposition state $|\Psi_N\rangle$ 
can be generated by applying an appropriate sequence \cite{Law} of evolution operators 
\begin{equation}
U^{(1)}_2(\tau^*_j)U^{(1)}_1(\tau_j){\ldots}U^{(1)}_2(\tau^*_1)U^{(1)}_1(\tau_1) 
\end{equation}
to the initial state $|g,0\rangle$,
where $\tau_i$ and $\tau^*_i$ are the $i$-th time steps for the time evolution operators $U^{(1)}_1$
and $U^{(1)}_2$, respectively, and $j$ is the total  number of steps.  
We assume that we can adjust the time duration of the interactions in each step.
The time steps $\tau_i$ and $\tau^*_i$ can be found inversely \cite{Law} by applying 
\begin{equation}
U^{(1)\dag}_1(\tau_1)U^{(1)\dag}_2(\tau^*_1){\ldots}U^{(1)\dag}_1(\tau_j)U^{(1)\dag}_2(\tau^*_j)
\end{equation}
to the state $|\Psi_N\rangle$ in Eq.~(\ref{multi-photon}).

We now consider two qubits (qubit 1 and qubit 2) for the quantum measurement.  We assume that
both qubits have the same form, Eq.~(\ref{Ham1}), of the interaction with the resonator.
Qubit 1 is used as an auxiliary qubit to disentangle qubit 2 from the resonator.
A CNOT gate $U_c$ can be applied to the two qubits. 
Details for implementing a CNOT gate can be found elsewhere (e.g., Refs.~\cite{Galiautdinov,Geller}).

Note that we now need to add a qubit-qubit coupling term in the Hamiltonian in Eq.~(\ref{Ham1}) (see red capacitor in Fig.~\ref{fig2}). 
The qubits in Fig.~\ref{fig1} can be coupled \cite{Martinis,McDermott,Matsuo} to each other via a capacitor so that ``spin-spin'' interactions 
can be produced.  This qubit-qubit coupling Hamiltonian, with the coupling strength $\lambda$, can be written as
\begin{eqnarray}
\label{qbcoup}
H_{\rm qb-qb}&=&\lambda(\sigma_{1+}\sigma_{2-}+\sigma_{1-}\sigma_{2+}).
\end{eqnarray}
Alternatively, the qubits can also be coupled via a high-excitation-energy quantum circuit such as a superconducting resonator \cite{Ashhab,DiCarlo,Ansmann}.
A CNOT gate can be realized, e.g., by either applying an external field \cite{Galiautdinov}, pulses \cite{Geller} or applying iSWAP gates [turning-on the qubit-qubit interaction in 
Eq.~(\ref{qbcoup}) for the time $t=\pi/4\lambda$] appropriately \cite{Bialczak}.  
By adjusting the frequencies of the two qubits, the CNOT gate and the qubit-resonator interaction
can be independently operated.
We choose qubit 1 as the control qubit and qubit 2 as the target qubit such that: 
$|gg\rangle{\rightarrow}|gg\rangle;~|ge\rangle{\rightarrow}|ge\rangle;~|eg\rangle{\rightarrow}|ee\rangle;~|ee\rangle{\rightarrow}|eg\rangle$.

Let us first prepare the resonator and the two qubits in a product state as
\begin{eqnarray}
|{\rm qubit1},{\rm qubit2},{\rm resonator}\rangle\!&=&\!|\Phi_N(0)\rangle,\\
\!&=&\!|gg\rangle\bigg(\frac{|0\rangle+|N\rangle}{\sqrt{2}}\bigg).
\end{eqnarray}
We let the resonator evolve freely for a time $\tau$, and
the state then becomes 
\begin{eqnarray}
|\Phi_N(\tau)\rangle&=&\exp(-i\omega{a^\dag{a}}\tau)|\Phi_N(0)\rangle,\\
&=&\frac{1}{\sqrt{2}}|gg\rangle\big[|0\rangle+\exp{(-i\varphi_N)}|N\rangle\big],
\end{eqnarray}
where $\varphi_N$ is the accumulated phase factor,
\begin{equation}
\varphi_N=\omega{N}\tau.
\end{equation}

Now we study how to transfer the relative phase between the two Fock states to the qubit 1.
We switch-on the interaction between the two qubits and the resonator sequentially, for the times $t^*_{N}$
and $t^*_{N-1}$ given by
\begin{equation}
t^*_N=\frac{\pi}{2{g_1}\sqrt{N}}~~{\rm and}~~t^*_{N-1}=\frac{\pi}{2{g_1}\sqrt{N-1}},
\end{equation}
where we set $g_1\approx{g_2}$. 
The state thus becomes
\begin{equation}
|\Phi_N(\tau_2)\rangle{~\approx~}\frac{1}{\sqrt{2}}\big[|gg\rangle|0\rangle-\exp{(-i\varphi_N)}|ee\rangle|N-2\rangle\big],
\end{equation}
where $\tau_2=\tau+t^*_N+t^*_{N-1}$.
We have assumed that the time duration $\tau$ is much greater than the times $t^*_{N}$ and $t^*_{N-1}$.
To perform the CNOT gate, we detune the qubits from the resonator and set the two qubits 
at the same frequency.  We then apply the CNOT gate such that the state $|ee\rangle$ will change to $|eg\rangle$.
The state then becomes
\begin{equation}
|\Phi_N(\tau_2')\rangle{~\approx~}\frac{1}{\sqrt{2}}\big[|gg\rangle|0\rangle-\exp{(-i\varphi_N)}|eg\rangle|N-2\rangle\big],
\end{equation}
where $\tau_2'=\tau_2+t_c$, and $t_c$ is the time duration for the CNOT gate.
We only turn-on the interaction between the qubit 2 and the resonator for 
a time $t^*_{N-2}=\pi/2{g_1}\sqrt{N-2}$, and this gives 
\begin{equation}
|\Phi_N(\tau_3)\rangle{\approx}\frac{1}{\sqrt{2}}\big[|gg\rangle|0\rangle+i\exp{(-i\varphi_N)}|ee\rangle|N-3\rangle\big],
\end{equation}
where $\tau_3=\tau_2'+t^*_{N-2}$.
In the interaction picture, we repeatedly apply the evolution operator $[U^{(2)}_2U_c]$ 
until the number of photons in the resonator becomes zero.  

We summarize this procedure in the interaction picture as 
\begin{widetext}
\begin{equation}
\label{dent_prod}
U_c\underbrace{U^{(2)}_2(t^*_1)U_c{\ldots\ldots}U^{(2)}_2(t^*_{N-2})U_c}_{2(N-2)~{\rm terms}}U^{(2)}_2(t^*_{N-1})U^{(1)}_2(t^*_{N})|\Phi_N(\tau)\rangle,
\end{equation}
\end{widetext}
where the time $t^*_{n}$ is
\begin{equation}
\label{t_n}
t^*_n=\frac{\pi}{2{g_1}\sqrt{n}},
\end{equation}
and $n$ is a positive number for $n=1,\ldots,{N}$.
Qubit 1 now completely disentangles from qubit 2 and the resonator. 
The final state becomes 
\begin{equation}
|\Phi_N(t_f)\rangle{\approx}\frac{1}{\sqrt{2}}\bigg\{|g\rangle+\exp{\bigg[-i\bigg(\varphi_N-\frac{3N\pi}{2}\bigg)\bigg]}|e\rangle\bigg\}|g\rangle|0\rangle,
\end{equation}
where $t_f=\tau+t_d$ and $t_d$ is the total time for disentanglement.
We assume that the free-evolution time $\tau$ is much larger than the time $t_d$ so that we
can ignore the relative phase accumulated during the time $t_d$.
Note that the relative phase $\varphi_N$ is encoded on the qubit 1.

Afterwards, we apply the $\pi/2$ pulse to the qubit 1 and then measure the excited state of qubit 1.
We can determine the phase factor from the probability $P_e$ of the excited state of qubit 1.
For simplicity, we set $\phi_1=0$ in Eq. (\ref{qubit_drive}).  
The probabilities of the excited state of qubit 1 then become
\begin{eqnarray}
P^{(1)}_e&\approx&\frac{1}{2}(1-\sin\varphi_N)~~{{\rm for}~N=4k},\nonumber\\
P^{(2)}_e&\approx&\frac{1}{2}(1+\cos\varphi_N)~~{{\rm for}~N=4k-1},\nonumber\\
P^{(3)}_e&\approx&\frac{1}{2}(1+\sin\varphi_N)~~{{\rm for}~N=4k-2},\nonumber\\
P^{(4)}_e&\approx&\frac{1}{2}(1-\cos\varphi_N)~~{{\rm for}~N=4k-3},\nonumber\\
\end{eqnarray}
where $k=1,\ldots,{N/4}$.

\subsubsection{Imperfect CNOT gate operations}
In realistic situations, the CNOT gate is not perfect due to decoherence or experimental constraints.
Let us briefly examine quantum phase measurements using imperfect CNOT gates.  Here we assume that
the fidelity of this imperfect CNOT gate is very high.
The small imperfections of this CNOT gate can be characterized by a parameter $\epsilon$ which is positive and 
close to one.  If this CNOT gate would be perfect, then the parameter $\epsilon$ would be equal to one.
A more detailed discussion of the effects of non-ideal CNOT gate operations on the disentanglement 
process is given in appendix B.  We repeat the same procedure in Eq.~(\ref{dent_prod}) using a number 
of imperfect CNOT gate operations.
The system can be described by the density matrix $\rho_{f}$
\begin{eqnarray}
\rho_f&\approx&\frac{\epsilon^{N-1}}{2}\big[|g\rangle\langle{g}|+|e\rangle\langle{e}|+\exp(i\varphi_N)|g\rangle\langle{e}|\nonumber\\
&&+\exp(-i\varphi_N)|e\rangle\langle{g}|\big]{\otimes}|g\rangle\langle{g}|0\rangle\langle{0}|,
\end{eqnarray}
where $\varphi_N=\omega{N}\tau$.
Here we have taken the leading order approximation of the density matrix (see appendix B).
We notice that qubit 1 cannot be fully disentangled from qubit 2 and the resonator
by using imperfect CNOT gates.
We now apply a $\pi/2$ pulse to qubit 1 and then measure the excited state of qubit 1.
For simplicity, we set $\phi_1=0$ in Eq.~(\ref{qubit_drive}).  
The probabilities of the excited states of qubit 1 are
\begin{eqnarray}
\label{prob_multi_1}
P^{(1)}_e&\approx&\frac{1}{2}\big(1-\epsilon^{N-1}\sin\varphi_N\big)~~~~{{\rm for}~N=4k},\nonumber\\
P^{(2)}_e&\approx&\frac{1}{2}\big(1+\epsilon^{N-1}\cos\varphi_N\big)~~~~{{\rm for}~N=4k-1},\nonumber\\
P^{(3)}_e&\approx&\frac{1}{2}\big(1+\epsilon^{N-1}\sin\varphi_N\big)~~~~{{\rm for}~N=4k-2},\nonumber\\
P^{(4)}_e&\approx&\frac{1}{2}\big(1-\epsilon^{N-1}\cos\varphi_N\big)~~~~{{\rm for}~N=4k-3},\nonumber\\
\end{eqnarray}
where $k=1,\ldots,{N/4}$.
The coherence factors, $\epsilon^{N-1}\sin\varphi_N$ or $\epsilon^{N-1}\cos\varphi_N$ 
in Eq.~(\ref{prob_multi_1}), contain a coefficient $\epsilon^{N-1}$ which is smaller than one.
This means that the imperfect CNOT gate  operations lead to dephasing of the qubit 1. 

\subsubsection{Dissipation in the photon field}
We now take into account the dissipation
effect of the photon field in Eq.~(\ref{master}) during the free evolution.  After a time $\tau$, the density matrix $\rho$ of 
the photon field can be written as \cite{Barnett}
\begin{eqnarray}
\rho&=&\frac{1}{2}\Bigg\{|0\rangle{\langle}0|+\exp{(i\varphi_N-\Gamma{N}{\tau})}|0\rangle\langle{N}|\nonumber\\
&&+\exp{(-i\varphi_N-\Gamma{N}{\tau})}|{N}\rangle\langle{0}|\nonumber\\
&&+\sum^{N}_{k=0}\frac{{N!}\exp{(-2k\Gamma{{\tau}})}[1-\exp{(-2\Gamma{{\tau}})}]^{N-k}}{k!(N-k)!}|k\rangle\langle{k}|\Bigg\}.\nonumber\\
\end{eqnarray}
We have also assumed that the decoherence of the qubit is negligible during the disentanglement process and the time $t_d$
is much smaller than the dissipation timescale $(\Gamma{N})^{-1}$.
The probabilities of the excited state of qubit 1 then become
\begin{eqnarray}
\label{prob_multi_2}
P^{(1)}_e&\approx&\frac{1}{2}(1+f_-)~~{{\rm for}~N=4k},\nonumber\\
P^{(2)}_e&\approx&\frac{1}{2}(1+h_+)~~{{\rm for}~N=4k-1},\nonumber\\
P^{(3)}_e&\approx&\frac{1}{2}(1+f_+)~~{{\rm for}~N=4k-2},\nonumber\\
P^{(4)}_e&\approx&\frac{1}{2}(1+h_-)~~{{\rm for}~N=4k-3},\nonumber\\
\end{eqnarray}
where $k=1,\ldots,{N/4}$, and $f_{\pm}$ and $h_{\pm}$ are two functions given by
\begin{eqnarray}
 f_{\pm}&=&{\pm}\epsilon^{N-1}\exp{(-\Gamma{N}{{\tau}})}\sin\varphi_N,\\
 h_{\pm}&=&{\pm}\epsilon^{N-1}\exp{(-\Gamma{N}{{\tau}})}\cos\varphi_N.
\end{eqnarray}

\subsubsection{Kerr nonlinearity}
We note that the superposition state in Eq.~(\ref{multi-photon}) can also be used to 
measure the phase due to the Kerr nonlinearity \cite{Gong,Semiao,Rebic}.  The Hamiltonian of the nonlinear interaction
is given by \cite{Rebic}
\begin{equation}
\label{nonlinear}
H_{\rm nonlinear}=\chi{a^{\dag{2}}}{a}^2,  
\end{equation}
where $\chi$ is the interaction strength.   
The two coupled CPB qubits can form a nonlinear medium of the photon field \cite{Rebic} (see also Ref.~\cite{Gong}).  
The strength $\chi$ of 
the nonlinear interaction in Eq.~(\ref{nonlinear}) can attain $1$ GHz or even higher \cite{Rebic}.
We only turn-on the nonlinear interaction during the free evolution.  Thus, the phase factor $\tilde{\varphi}_N$
can be rewritten as 
\begin{equation}
\tilde{\varphi}_N{~=~}\tilde{\omega}N{\tau},
\end{equation}
where 
\begin{equation}
\label{effomega}
\tilde{\omega}=\omega+\chi(N-1)
\end{equation}
is an effective frequency due to the Kerr nonlinearity.
We then apply the same procedure to
detect the phase of the photon field. 

\subsubsection{Visibility}
Now we investigate the visibility of the quantum coherence.  
The visibilities $V^{(1)}$ and $V^{(2)}$ denote the 
odd and even photon numbers, respectively.  
From Eqs.~(\ref{visibility}) and (\ref{prob_multi_2}), the visibility $V^{(1)}$ for an odd number of photons can be found as
\begin{eqnarray}
V^{(1)}&=&\frac{\epsilon^{N-1}}{2}\bigg[1+\bigg(\frac{\tilde{\omega}^2}{\tilde{\omega}^2+{\Gamma}^2}{\bigg)}^{\frac{1}{2}}\exp{(-{\Gamma}{\tau_{1})}}\bigg],
\\
\tau_{1}&=&\frac{1}{\tilde{\omega}}\bigg\{\arccos\bigg[-\bigg(1+\frac{{\Gamma}^2}{\tilde{\omega}^2}\bigg)^{-\frac{1}{2}}\bigg]\bigg\},
\end{eqnarray}
where $\epsilon{~\leq~}1$ is the parameter that characterizes the quality of the imperfect CNOT gate 
and $\tilde{\omega}=\omega+\chi{(N-1)}$ is the effective frequency in Eq.~(\ref{effomega}).
  
For an even number of photons,
we take the form of the coherence factor 
\begin{equation}
C(\tau)=\epsilon^{N-1}\exp{(-{\Gamma}N\tau)}\sin\tilde{\omega}\tau.  
\end{equation}
The visibility $V^{(2)}$ for even photon numbers can be found as ($\tilde{\omega}>{\Gamma}$)
\begin{eqnarray}
V^{(2)}\ &{\approx}&\ \frac{\epsilon^{N-1}}{2}\bigg[\exp\bigg({-\frac{\pi{\Gamma}}{2\tilde{\omega}}}\bigg)+\exp\bigg({-\frac{3\pi{\Gamma}}{2\tilde{\omega}}}\bigg)\bigg].
\end{eqnarray}
The visibility $V^{(1)}$ is slightly higher than $V^{(2)}$.

For $\chi=0$ and $\epsilon~=~1$, we find that the visibilities have no difference between 
the cases of single-photon state $|\Psi_1\rangle$ and 
multi-photon superposition state $|\Psi_N\rangle$ in Eq.~(\ref{multi-photon}).  
The degree of visibility depends on the ratio of $\omega$ and $\Gamma$.  
However, the visibilities of the superposition of multi-photon states can be greatly 
enhanced with the nonlinear interaction strength $\chi$.   
For $\epsilon=1$, the degree of visibility depends 
on the ratio $\tilde{\omega}/\Gamma=(\omega+\chi{N})/\Gamma$ from Eq.~(\ref{effomega}) 
such that higher visibilities can be obtained for larger number of photons ($N\gg{1}$).  
This means that the superposition of multi-photon states can 
show a high contrast of interference fringes, even in the presence of dissipation.

\section{Factorizing integers using superconducting circuits and Gauss sums}
We have shown that the quantum phase of the photon states in a superconducting resonator can be measured 
with qubits.  These measurement methods are useful for the implementation of the Gauss sum (this section)
and metrology (the following section).

Now we study a physical realization of the Gauss sum algorithm using superconducting circuits.  
The Gauss sum can verify whether a number $\tilde{n}$ is a factor of another number $\tilde{N}$ 
or not.  The Gauss sum \cite{Mack,Mehring} is defined by
\begin{eqnarray}
\label{Gauss_sum}
C_{\tilde{N}}(\tilde{n})&=&\frac{1}{\tilde{n}}\sum^{\tilde{n}-1}_{k=0}\exp\bigg(-2\pi{i}k^2\frac{\tilde{N}}{\tilde{n}}\bigg).
\end{eqnarray}
If $\tilde{n}$ is a factor of $\tilde{N}$, then $|C_{\tilde{N}}(\tilde{n})|$ is equal to one.
Otherwise, the value of $|C_{\tilde{N}}(\tilde{n})|$ is less than one.   
To find the factors $\tilde{n}$ of the integer $\tilde{N}$,
the trial factor $\tilde{n}$ scans through all numbers from 2 to $\sqrt{\tilde{N}}$ \cite{Mehring,Gilowski,Bigourd}.   
Thus, in this manner, it can factorize integers.  However, this Gauss sum algorithm does not provide
a speed-up over classical computation. 

To save considerable experimental resources and minimize the effects of decoherence,
it is advantageous to use as few terms as possible in the Gauss sum in Eq.~(\ref{Gauss_sum}) \cite{Gilowski}.
Thus, the truncated Gauss sum can be employed \cite{Gilowski},
\begin{eqnarray}
\label{tGS}
C^{K}_{\tilde{N}}(\tilde{n})&=&\frac{1}{K+1}\sum^{K}_{k=0}\exp\bigg(-2\pi{i}k^2\frac{\tilde{N}}{\tilde{n}}\bigg),
\end{eqnarray}
where $K$ is a positive integer which is smaller than $\tilde{n}$.  
We only need to sum over $K+1$ terms instead of the total $\tilde{n}$ 
terms so that considerable experimental resources can be saved.

\subsection{Single-photon case}
We can apply the same method in determining the phase factor of the photon field in the resonator  
using the superposition of the vacuum and single-photon state in Eq.~(\ref{multi-photon}).  
We follow the same method in Sec. III.A 
to generate the superposition of the vacuum and the single-photon state 
with equal weights, i.e., 
\begin{equation}
|\tilde{\Phi}_1\rangle=\frac{1}{\sqrt{2}}|g\rangle\big(|0\rangle+|1\rangle\big).
\end{equation}
Waiting the following times
\begin{equation} 
\tau_k=2{\pi}k^2\frac{\tilde{N}}{\tilde{n}\omega},
\end{equation}
for the free-evolution,  allows the state to accumulate a relative phase $(-i\omega{\tau_k})$
between the two Fock states $|0\rangle$ and $|1\rangle$.
The state then becomes
\begin{equation}
|\tilde{\Phi}_1(\tau_k)\rangle=\frac{1}{\sqrt{2}}|g\rangle\big[|0\rangle+\exp(-i\omega\tau_k)|1\rangle\big].
\end{equation}
Using the method described in Sec. III.A, we can transfer the relative phase to qubit 1.
The state of qubit 1 should then be measured after applying a $\pi/2$-pulse to it.
The phase factor $\cos(\omega\tau_k)$ can be determined from the probability of the excited state of 
qubit 1 in Eq.~(\ref{prob_e}).  

By repeating the same procedure $K+1$ times and setting the time duration $\tau_k$ for $k=0,\ldots,K$,
we take the average of the total sum.
We readily obtain the real part of the truncated Gauss sum in Eq.~(\ref{tGS}) as \cite{Gilowski}
\begin{equation}
\label{realGsum}
R^{K}_{\tilde{N}}(\tilde{n})=\frac{1}{K+1}\sum^{K}_{k=0}\cos\bigg(2\pi{k^2}\frac{\tilde{N}}{\tilde{n}}\bigg).
\end{equation}
Indeed, the real part of the truncated Gauss sums has been shown to experimentally find the factors of integers \cite{Gilowski}.

\begin{figure}[ht]
\centering
\includegraphics[height=8cm]{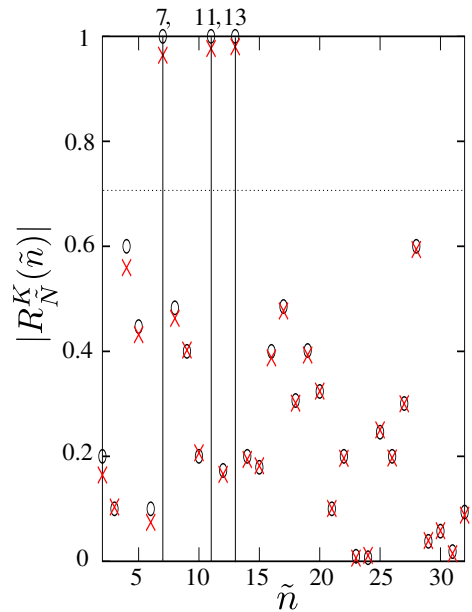}
\caption{ \label{fig3} (Color online) Truncated Gauss sum for $\tilde{N}=1001=7\times{11}\times{13}$, 
plotted versus the trial factors $\tilde{n}$.
Results for a damping rate $\Gamma=0$ (no dissipation) 
are shown with black circles; while 
$\Gamma=6.92\times{10^{-6}}\omega$ are shown with red crosses.  
We have summed over $K+1=5$ terms for $\tilde{n}=2,{\ldots},\sqrt{1001}{~\approx~}{32}$. 
The factors of $\tilde{N}$ are marked by the vertical lines located at $\tilde{n}=7,11$ and 13.   
All factors are above the threshold $1/\sqrt{2}\approx0.7071$, 
indicated by a thin horizontal dotted line.
}
\end{figure}

Now we study how to apply the truncated Gauss sum for checking factors.  In Fig. \ref{fig3}, 
the truncated Gauss sum $|R^{K}_{\tilde{N}}(\tilde{n})|$ for $\tilde{N}=1001=7\times{11}\times{13}$ 
is plotted against 
the trial factors $\tilde{n}$.  The truncated Gauss sums are represented by black circles.
We have only used $K+1=5$ terms in the truncated Gauss sum for $\tilde{n}$ by 
scanning $\tilde{n}$ from 2 to $\sqrt{1001}{~\approx~}{31}$. 

In Ref. \cite{Stefanak}, \v{S}tefa\v{n}\'{a}k {\it et al.} gave 
a threshold to discriminate factors from non-factors.   They found that 
the truncated Gauss sum for non-factors is bounded from above by $1/\sqrt{2}$ in the limit of large
$K$ \cite{Stefanak}.
As shown in Fig. \ref{fig3}, we can see that all factors are above the threshold of the Gauss sums, 
$1/\sqrt{2}\approx{0.7071}$ \cite{Stefanak}, whereas all non-factors are below the threshold, 
$1/\sqrt{2}\approx{0.7071}$, shown by the horizontal dotted line.  
Therefore, this enables us to clearly distinguish the factors from nonfactors.

\subsubsection{Effect of dissipation of the photon field}
We now examine the truncated Gauss sum in the presence of dissipation of the photon field. 
We adopt the same method as discussed in the previous subsection to determine the phase
factor.  From Eq.~(\ref{P_es}), the truncated Gauss sums can be written as
\begin{equation}
\label{tGS1}
R^{K}_{\tilde{N}}(\tilde{n})=\frac{1}{K+1}\sum^{K}_{k=0}\exp{(-\Gamma{\tau_k})}\cos\bigg(2\pi{k^2}\frac{\tilde{N}}{\tilde{n}}\bigg).
\end{equation}
The performance of the Gauss sums can decrease due to dissipation.  
The terms with higher $k$ can become vanishingly small in Eq.~(\ref{tGS1}).
This limits the size of the number $\tilde{N}$ to be vertified by the Gauss sum.  
In Fig. \ref{fig3}, we plot the 
truncated Gauss sum $R^{K}_{\tilde{N}}(\tilde{n})$ as a function of the trial
numbers $\tilde{n}$ for $\tilde{N}=1001=7\times{11}\times{13}$.
Here we consider the damping rate $\Gamma/\omega=6.92\times{10^{-6}}$ 
(we have taken the values of $\omega/2\pi=6.57$~GHz and $\Gamma=2.86\times{10^5}$~Hz from 
the experiment in \cite{Hofheinz1}). 
We then sum over $K+1=5$ terms for checking the 31 numbers $\tilde{n}=2,\ldots,{\sqrt{1001}}{~\approx~}{32}$.  
As shown in Fig. \ref{fig3}, the truncated Gauss sums slightly
decrease due to damping (denoted by the red cross).  
However, we can see that the truncated Gauss sum can still be used 
for distinguishing the factors from nonfactors if the relaxation 
times of resonators are about several $\mu$s \cite{Hofheinz1}.


\subsection{Multi-photon case}
\begin{figure}[ht]
\centering
\includegraphics[height=8cm]{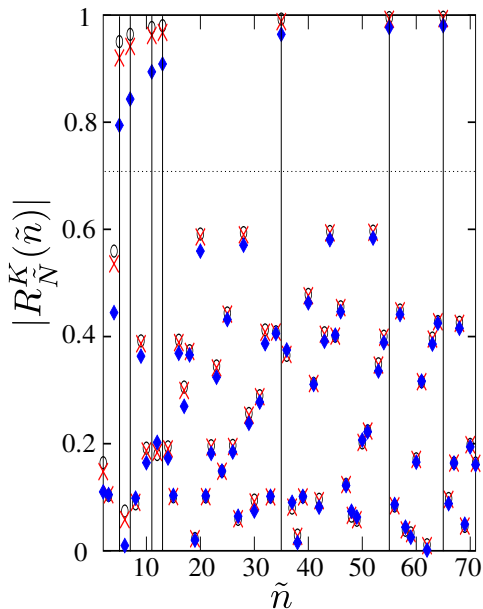}
\caption{ \label{fig4} (Color online) The truncated Gauss sum with $K+1=5$ terms 
for $\tilde{N}=5005=5\times{7}\times{11}\times{13}$ 
is plotted versus the trial factors $\tilde{n}$, for the nonlinear Kerr interaction strength $\chi=\omega$ 
and damping rate $\Gamma=6.92\times{10}^{-6}\omega$.
The blue diamonds, red crosses and black circles represent the different numbers of photons: $N=1$, 3 and 5,
respectively.
The factors of $\tilde{N}$ (here: 5, 7, 11, 13, 35, 55 and 65) 
are indicated by the vertical lines.   All factors are above the threshold $1/\sqrt{2}\approx0.7071$, 
indicated by a thin horizontal dotted line.   
}
\end{figure}

Next we consider the multi-photon superposition states $|\Psi_N\rangle$ in Eq.~(\ref{multi-photon}) for 
the truncated Gauss sum.   We first produce the superposition of the vacuum and the multi-photon state 
with equal weights, i.e., 
\begin{equation}
|\tilde{\Phi}_N\rangle=\frac{1}{\sqrt{2}}|gg\rangle\big(|0\rangle+|N\rangle\big).
\end{equation}
We let the system freely evolve for the times
\begin{equation} 
\label{tau}
\tilde{\tau}_k=2{\pi}k^2\frac{\tilde{N}}{\tilde{n}\tilde{\omega}{N}},
\end{equation}
where $\tilde{\omega}=\omega+\chi(N-1)$ is the effective frequency in Eq.~(\ref{effomega})
if the Kerr nonlinearity is used.    
Then, a relative phase between the two Fock states is acquired.
The state becomes 
\begin{equation}
|\tilde{\Phi}_N(\tilde{\tau}_k)\rangle=\frac{1}{\sqrt{2}}\big[|0\rangle+\exp(-i\tilde{\omega}\tilde{\tau}_k)|N\rangle\big].
\end{equation}
We now apply the phase detection method described in Sec. III.B for the multi-photon case.  
After applying a $\pi/2$-pulse to qubit 1, we measure the state of qubit 1.  
Then, for an odd number of photons, we can determine the phase factor $\cos(\omega\tau_k)$ from 
the probabilities of qubit 1 in Eq.~(\ref{prob_multi_1}).  
The truncated Gauss sum is given by
\begin{eqnarray}
R^{K}_{\tilde{N}}(\tilde{n})=\frac{\epsilon^{N-1}}{K+1}\sum^{K}_{k=0}\cos\bigg(2\pi{k^2}\frac{\tilde{N}}{\tilde{n}}\bigg),
\end{eqnarray}
where $\epsilon^{N-1}$ is a parameter originating from imperfect CNOT gate operations.
In the presence of dissipation of the photon field,
we can determine the phase factor from the probabilities in Eq.~(\ref{prob_multi_2}).
The truncated Gauss sum becomes
\begin{equation}
\label{tGS2}
R^{K}_{\tilde{N}}(\tilde{n})=\frac{\epsilon^{N-1}}{K+1}\sum^{K}_{k=0}\exp{(-{\Gamma}{N}{\tilde{\tau}_k})}\cos\bigg(2\pi{k^2}\frac{\tilde{N}}{\tilde{n}}\bigg).
\end{equation}

\subsection{Kerr nonlinearity}
Now we study the performance of 
the truncated Gauss sum for checking factors using the Kerr nonlinearity.
In Fig. \ref{fig4}, the truncated Gauss sum is plotted versus the trial factors $\tilde{n}$
for $\tilde{N}=5005=5\times{7}\times{11}\times{13}$.
The truncated Gauss sums with various numbers of photons [$N=1$ (blue diamonds), $N=3$ (red crosses)
and $N=5$ (black circles)] are shown in Fig. \ref{fig4}.  
We can see that the truncated Gauss sum for the factors 
are much closer to unity if the larger photon numbers are used.
The truncated Gauss sum using the Kerr nonlinearity can show a clearer pattern to discriminate 
the factors from nonfactors and a much larger number can be verified \cite{Miranowicz}.
This can be easily understood by rewriting 
the truncated Gauss sum [from Eqs.~(\ref{tau}) and (\ref{tGS2})] as
\begin{equation}
\label{tGS3}
R^{K}_{\tilde{N}}(\tilde{n})=\frac{1}{K+1}\sum^{K}_{k=0}\exp\!{\bigg(\!-\frac{2{\pi}k^2{\Gamma}\tilde{N}}{\tilde{\omega}\tilde{n}}\bigg)}\cos\bigg(2\pi{k^2}\frac{\tilde{N}}{\tilde{n}}\bigg).
\end{equation}
The exponential functions in Eq.~(\ref{tGS3}) are closer to one when the effective frequency 
$\tilde{\omega}=\omega+\chi{(N-1)}$ becomes higher.
The larger number of photons leads to a higher value of $\tilde{\omega}$.
Therefore, the use of Kerr nonlinearities enhances the performance of the truncated Gauss sum for finding the
factors of an integer, even in the presence of dissipation in the photon field.

We can roughly estimate the size of the factorized number $\tilde{N}$ for 
which the exponential factor in Eq.~(\ref{tGS3}) is around $\exp{(-1)}$.   
In this case, the fully factorizable numbers are then multiplied by a factor of $N$ and 
it can attain $10^2{N}$ if the nonlinear strength $\chi$ is about
the frequency of the resonator, and with the same values for the parameters discussed above and $K\lesssim{10}$. 
If the relaxation time of the resonator were to be $\sim~1~\mu$s, $\sim~{1}~$ ms and $\sim~{1}~$ s, 
then the largest partially-factorizable numbers would be $\sim~10^4{N}$, $\sim{10^7}N$  and $\sim{10^{10}}N$, 
respectively, where $N$ is the number of photons in the resonator.

\subsubsection{Effect of imperfect CNOT gate operations}
\begin{figure}[ht]
\centering
\includegraphics[height=8.0cm]{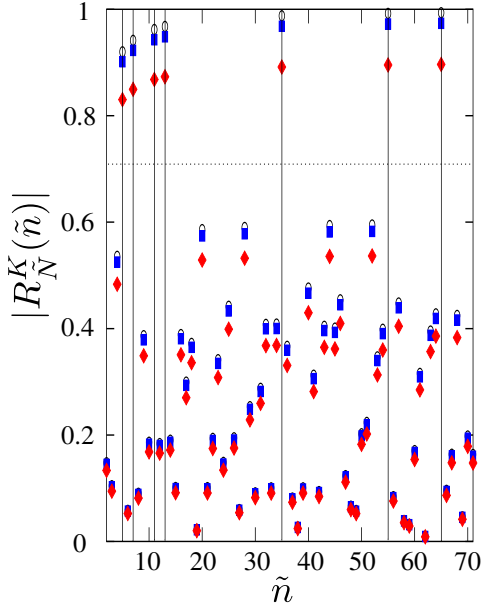}
\caption{ \label{fig5} (Color online) Truncated Gauss sum with $K+1=5$ terms for $\tilde{N}=5005=5\times{7}\times{11}\times{13}$ 
plotted versus the trial factors $\tilde{n}$, where the nonlinear Kerr interaction strength 
$\chi=\omega$, the damping rate $\Gamma=6.92\times{10}^{-6}\omega$, 
and the number of photons $N$ is 3.
Different values of $\epsilon$ are shown: $\epsilon=1$ (black circles), $\epsilon=0.99$ (blue squares)
and $\epsilon=0.95$ (red diamonds).
The factors of $\tilde{N}$ are indicated by the vertical lines.   All factors are above the threshold $1/\sqrt{2}\approx0.7071$, 
indicated by a thin horizontal dotted line. 
}
\end{figure}

Note that we require more CNOT gate  operations to disentangle qubit 1 from both qubit 2 and the resonator.
This requirement is for the multi-photon states involving higher number of photons in Eq.~(\ref{multi-photon}).  
From Eq.~(\ref{tGS2}), the truncated Gauss sum scales with the coefficient $\epsilon^{N-1}$.
Thus, the imperfect CNOT gates unavoidably affect
the performance of the truncated Gauss sum.   
  
In Fig. \ref{fig5}, we plot the
truncated Gauss sum for $\tilde{N}=5005=5\times{7}\times{11}\times{13}$ 
against the trial factors $\tilde{n}$,  using the multi-photon superposition state for 
$N=3$ in Eq.~(\ref{multi-photon}). 
We find that the performance of the truncated Gauss sum is very sensitive to small variations of $\epsilon$. 
As shown in Fig. \ref{fig5}, the values of the truncated Gauss sum for $\epsilon=0.95$ (red diamonds) are much lower 
than the case for $\epsilon=0.99$ (blue squares) and $\epsilon=1$ (black circles).
This may limit the use of multi-photon states $|\Psi_N\rangle$ for the truncated Gauss sum
if the photon number $N$ becomes large.


\section{Precision measurement of the resonator's frequency}
We can determine the quantum phase factor of the photon states by detecting the state of the qubit.
This should enable to precisely measure the frequency of the resonator $\omega$. 
The uncertainty $|\delta{\omega}|$ of the frequency $\omega$ of the resonator is given by \cite{Huelga,Braunstein1,Braunstein2}
\begin{eqnarray}
|\delta{\omega}|&=&\frac{1}{\sqrt{M}}\frac{\sqrt{P_e(1-P_e)}}{\bigg|\dfrac{dP_e}{d\omega}\bigg|},
\end{eqnarray}
where $M$ is the number of measurements.  

\subsection{Single-photon case}
We now consider the detection with the superposition of the vacuum and the single-photon
state. Using Eq.~(\ref{prob_e}), the minimum value of the uncertainty $|\delta{\omega}|_{\rm min}$,  for $\Gamma=0$, is
\begin{eqnarray}
|\delta{\omega}|_{\rm min}=\frac{1}{\sqrt{M}\tau},
\end{eqnarray}
where $\tau=m\pi/2\omega$ is the time duration of each measurement and $m$ is an odd integer.  

In the presence of dissipation ($\Gamma{\neq}0$), the uncertainty is given by
\begin{eqnarray}
|\delta{\omega}|&=&\bigg[{\frac{1-\exp{(-2\Gamma{\tau})}\cos^2(\omega{\tau})}{M{\tau^2}\exp{(-2\Gamma{\tau})}\sin^2(\omega\tau)}}\bigg]^{{1}/{2}}.
\end{eqnarray}
We assume that the frequency $\omega$ is much greater than the damping rate $\Gamma$.
The minimum uncertainty $|\delta\omega|_{\rm min}$ can be found as 
\begin{equation}
|\delta{\omega}|_{\rm min}=\frac{1}{\sqrt{M}}{\exp(1)\Gamma}
\end{equation} 
for $\omega\tau=m\pi/2$ and 
\begin{equation}
\tau=\frac{1}{\Gamma}.
\end{equation}

The precise measurement of the resonator's frequency $\omega$ can be
used to determine the length of the resonator. The frequency of the resonator $\omega$ 
is proportional to ${\pi{c}/L}$, where
$c$ is the speed of light and $L$ is the length of resonator.  This enables us to 
measure the macroscopic quantity $L$ precisely and its accuracy is up to 
\begin{eqnarray}
|\delta{L}|=\frac{L|\delta{\omega}|}{\omega}.
\end{eqnarray}

\subsection{Multi-photon case}
Next we study the degree of accuracy in the phase measurement with the superposition
of the multi-photon state $|\Psi_N\rangle$ in Eq.~(\ref{multi-photon}). 
Without dissipation ($\Gamma=0$), the minimum value of the uncertainty $|\delta{\omega}|_{\rm min}$ of the frequency $\omega$ is
\begin{equation}
|\delta{\omega}|_{\rm min}{~\approx~}\frac{1}{\sqrt{M}}\frac{\epsilon^{1-N}}{N\tau}.
\end{equation}
\vspace{.2cm}
This accuracy scales as $\epsilon^{1-N}/{N}$ with the superposition state $|\Psi_N\rangle$ which gives a much better improvement than 
using the single-photon state $|\Psi_1\rangle$ if $\epsilon$ is very close to one.
However, in the presence of dissipation ($\Gamma{\neq}0$), the minimum uncertainty of $|\delta{\omega}|_{\rm min}$ becomes 
\begin{eqnarray}
|\delta{\omega}|_{\rm min}\approx\frac{1}{\sqrt{M}}{\exp(1)\epsilon^{1-N}\Gamma}.
\end{eqnarray}
It is no different to the single-photon case even if $\epsilon=1$.

However, the nonlinear interaction strength $\chi$ can be measured precisely.
The minimum uncertainty $|\delta{\chi}|_{\rm min}$ is given by
\begin{equation}
|\delta{\chi}|_{\rm min}=\frac{1}{\sqrt{M}}\frac{\exp(1)\epsilon^{1-N}\Gamma}{N}.
\end{equation}
This uncertainty scales with $\epsilon^{1-N}/N$ under decoherence.   This enables to 
detect the strength of the Kerr nonlinearity with very high accuracy if $\epsilon^{1-N}/N$
is very small.

\vspace*{.2cm}
\section{Summary}
\vspace*{.2in}
We have presented a method to measure the quantum coherence of the superposition of
two number states in a superconducting resonator with two Josephson phase qubits.  
We have also studied the visibility of the quantum coherence in a dissipative environment.
We found that the visibility of the photon can be enhanced 
if nonlinear interactions are used.  This may be useful to probe the quantum coherence
of multi-photon superposition states.  We showed that the phase measurement scheme can
be applied to factorizing integers and parameter estimation.

The detection of the superposition of the vacuum and the single-photon state
can be realized with current technology \cite{Hofheinz1}. 
But the measurement of multi-photon superposition states 
involves a number of CNOT gate operations
to disentangle the qubit from the resonator.   
The quality of measurements is degraded due to the imperfect operations of the CNOT gate.
Also, significant resources can be consumed when measuring the superposition
of multi-photon states because the number of gate operations is proportional to 
the number of photons $N$ being detected.  Here we proposed a method which can be used to detect
superposition states with a few photons.

\vspace*{0.2cm}
\begin{acknowledgments}
\vspace*{0.2cm}
We thank Sahel Ashhab, Max Hofheinz and Adam Miranowicz for useful
discussions and comments.  FN acknowledges partial support from the 
Laboratory of Physical Sciences, 
National Security Agency, Army Research Office, DARPA,
National Science Foundation grant No.~0726909,
JSPS-RFBR contract No.~09-02-92114, 
Grant-in-Aid for Scientific Research (S), 
MEXT Kakenhi on Quantum Cybernetics, and 
Funding Program for Innovative R\&D on S\&T (FIRST).
\end{acknowledgments}

\appendix
\section{Qubit-resonator coupling in the far-detuning regime}
The Hamiltonian $H^{(j)}=H_{\rm res}+H^{(j)}_{\rm qbit}+H^{(j)}_{\rm qbit-res}$
\begin{eqnarray}
H^{(j)}&=&\omega{a^\dag{a}}+\frac{\omega_{0j}}{2}\sigma_{jz}+g_j(a\sigma_{j+}+\sigma_{j-}a^\dag)
\end{eqnarray}
can be exactly solved \cite{Scully} for $j=1,2$.  We now 
consider the subspace spanned by the basis $|g\rangle|n\rangle$ and $|e\rangle|n-1\rangle$.
The two eigenvalues can be solved as 
\begin{eqnarray}
\lambda^{(j)}_{\pm}&=&\omega\Big(n-\frac{1}{2}\Big)\pm\frac{\delta_j}{2},
\end{eqnarray}
and the corresponding eigenvectors are 
\begin{eqnarray}
\label{eigvec1}
|+\rangle_j&=&\sin\beta_j|g\rangle|n\rangle+\cos\beta_j|e\rangle|n-1\rangle,\\
\label{eigvec2}
|-\rangle_j&=&\cos\beta_j|g\rangle|n\rangle-\sin\beta_j|e\rangle|n-1\rangle,
\end{eqnarray}
where 
\begin{eqnarray}
\delta_j&=&\sqrt{\Delta^2_j+4g^2_j{n}},\\
\sin\beta_j&=&-\frac{\Delta_j+\delta_j}{\sqrt{(\Delta_j+\delta_j)^2+4g^2_jn}},\\
\cos\beta_j&=&\frac{2g_j\sqrt{n}}{\sqrt{(\Delta_j+\delta_j)^2+4g^2_jn}},\\
\Delta_j&=&\omega_{0j}-\omega.
\end{eqnarray}

If the detuning $\Delta_j=\omega_{0j}-\omega$ is much larger than the interaction strength $g_j$,
the two eigenvectors becomes 
\begin{equation}
|+\rangle_j\approx|e\rangle|n-1\rangle,~~~~|-\rangle_j\approx|g\rangle|n\rangle,
\end{equation}
in Eqs.~(\ref{eigvec1}) and (\ref{eigvec2}),
respectively.  Therefore, the interaction between the qubit and the photon field can be 
effectively turned-off by far-detuning the qubit from the resonator.

\section{Analysis of imperfect CNOT gate operations on the disentanglement process}
In this appendix, we study the effects of imperfect CNOT gate operations on the disentanglement process.   
We consider the CNOT gate operations to be described by a process which suffers from dissipation, decoherence
and imperfections \cite{Poyatos}.  In general, this operation can be represented by $\mathcal{E}$ \cite{Poyatos} such that 
\begin{equation}
\rho_{\rm out}=\mathcal{E}(\rho_{\rm in}),
\end{equation}
where $\rho_{\rm in}$ and $\rho_{\rm out}$ are the density matrices of the 
input and output states, respectively.  The operation $\mathcal{E}$ is a convex-linear map
and also a positive map \cite{Nielsen}.  Here we assume the operation $\mathcal{E}$ is trace-preserving, i.e.,
${\rm tr}[\mathcal{E}(\rho_{in})]=1$.

It is convenient to write the states as
\begin{equation}
\begin{split}
|0\rangle&=|gg\rangle,~~~~|1\rangle=|ge\rangle,\\
|2\rangle&=|ee\rangle,~~~~|3\rangle=|eg\rangle.
\end{split}
\end{equation}
An ideal CNOT gate  is defined as
\begin{equation}
\begin{split}
|0\rangle&{\rightarrow}|0\rangle,~~~~|1\rangle\rightarrow|1\rangle,\\
|2\rangle&{\rightarrow}|3\rangle,~~~~|3\rangle\rightarrow|2\rangle.
\end{split}
\end{equation}
Now we consider the non-ideal CNOT gate operation but with a high fidelity as:
\begin{eqnarray}
\langle{0}|\mathcal{E}(|0\rangle\langle{0}|)|0\rangle&\approx&\langle{1}|\mathcal{E}(|1\rangle\langle{1}|)|1\rangle\approx{\epsilon},\\
\langle{3}|\mathcal{E}(|2\rangle\langle{2}|)|3\rangle&\approx&\langle{2}|\mathcal{E}(|3\rangle\langle{3}|)|2\rangle\approx{\epsilon},\\
\langle{0}|\mathcal{E}(|0\rangle\langle{1}|)|1\rangle&\approx&\langle{0}|\mathcal{E}(|0\rangle\langle{2}|)|3\rangle\approx{\epsilon},\\
\langle{0}|\mathcal{E}(|0\rangle\langle{3}|)|2\rangle&\approx&\langle{1}|\mathcal{E}(|1\rangle\langle{2}|)|3\rangle\approx{\epsilon},\\
\langle{1}|\mathcal{E}(|1\rangle\langle{3}|)|2\rangle&\approx&{\epsilon},
\end{eqnarray}
and $\mathcal{E}(|i\rangle\langle{j}|)^*=\mathcal{E}(|j\rangle\langle{i}|)$,
where $\epsilon$ is a positive number close to one.   Otherwise, 
the remaining terms
$\langle{i'}|\mathcal{E}(|i\rangle\langle{j}|)|j'\rangle$ are
equal to small parameters $\epsilon^{i'j'}_{ij}$, which are much smaller than $\epsilon$, where $i,i',j$ and $j'=0,1,2,{3}$.

We now follow the same disentanglement procedure as summarized in Eq.~(\ref{dent_prod}), by using the imperfect CNOT gates.
The density matrix of the total system becomes
\begin{eqnarray}
\label{npfstate}
\rho_f&=&\frac{\epsilon^{N-1}}{2}\big[|g\rangle\langle{g}|+|e\rangle\langle{e}|+\exp(i\varphi_N)|g\rangle\langle{e}|\nonumber\\
&&+\exp(-i\varphi_N)|e\rangle\langle{g}|\big]{\otimes}|g\rangle\langle{g}|0\rangle\langle{0}|+\mathcal{O}\big(\epsilon^{N-2}\epsilon^{i'j'}_{ij}\big),\nonumber\\
\end{eqnarray}
where $\varphi_N=\omega{N}{\tau}$.
The leading order approximation is the density matrix containing the first four terms with the coefficients $\epsilon^{N-1}$.  
The remaining terms involving a large number of entangled states of qubit 1, qubit 2 and the resonator
are of the order of $\epsilon^{N-2}\epsilon^{i'j'}_{ij}$. 
Using imperfect CNOT gates, qubit 1 cannot be completely disentangled from qubit 2 and the resonator in Eq. (\ref{npfstate}).

\end{document}